\documentclass[a4paper,11pt]{article}

\usepackage[T1]{fontenc}
\usepackage[utf8]{inputenc}
\usepackage[section]{placeins}
\usepackage{cite}

\usepackage{geometry}
\usepackage{amsmath,amssymb,bm}
\usepackage{graphicx}
\usepackage{hyperref}
\usepackage{tikz}
\usetikzlibrary{positioning}

\title{\textbf{Dark matter relic abundance from a critical-density instability}}

\author{Hindi Zouhair$^{a}$\thanks{\texttt{Hindizouhair@gmail.com}},\hspace{0.5cm}
	\\[6pt]
	$^{a}$\textit{Laboratory of High Energy Physics (LHEP-MS), Mohammed V University, Rabat, Morocco
	}}

\date{}
\begin{document}
	\maketitle

\begin{abstract}
	We study a nonstandard dark-matter thermal history in which strong self-interactions give rise to collective many-body effects at high number density, as in strongly interacting quantum media. At early times, dark matter occupies a correlated phase in which its coupling to a light mediator is dynamically screened, suppressing annihilation far below the
	perturbative rate.	As the Universe expands and the number density decreases, this screened phase becomes unstable at a critical density $n_c$, triggering a rapid, far-from-equilibrium annihilation episode.
	We show that this annihilation burst fixes the final relic abundance, which is governed	primarily by $n_c$ rather than by the microscopic annihilation coupling. Using a minimal effective parametrization, we solve the resulting modified Boltzmann 	evolution and map the viable parameter space.
	For TeV-scale dark matter and sub-GeV mediators, we find relic abundances consistent with observations together with self-interaction cross sections relevant for small-scale structure, realizing a consistent and predictive nonstandard thermal history.
\end{abstract}


\section{Introduction}

The existence of dark matter (DM) is firmly established by a wide range of
astrophysical and cosmological observations.
In particular, the cosmic microwave background and large-scale structure
measurements determine the present-day DM relic abundance to be
$\Omega_{\rm DM} h^2 \simeq 0.12$ with percent-level precision~\cite{Planck:2018vyg,Bertone:2005}.
Understanding the origin of this value remains one of the central challenges
in particle cosmology.

In the standard weakly interacting massive particle (WIMP) paradigm,
the observed relic abundance arises from thermal freeze-out of a stable
particle species with weak-scale mass and interactions~\cite{Lee:1977ua,Kolb:EarlyUniverse,Griest:1990kh,GondoloGelmini:1991,Jungman:1996}.

In this framework, the DM abundance is determined by the annihilation
cross section through the freeze-out condition
$n_\chi \langle\sigma v\rangle \sim H$.
While this mechanism remains theoretically appealing, the absence of
experimental signals has motivated extensive exploration of alternative
production mechanisms and nonstandard thermal histories~\cite{GelminiGondolo:2006}.

A broad class of scenarios has been proposed, including freeze-in of feebly
interacting massive particles~\cite{Hall:2009bx,Bernal:2017}, asymmetric dark matter~\cite{PetrakiVolkas:2013,Zurek:2014},
and hidden sectors with nontrivial internal dynamics such as cannibal dark
matter~\cite{Pappadopulo:2016pkp,Farina:2016llk}.
Independently, self-interacting dark matter (SIDM) has attracted significant
interest as a possible solution to small-scale structure anomalies, requiring
sizeable DM self-scattering cross sections~\cite{Spergel:1999mh}.
These developments motivate the study of dark sectors in which collective
effects and many-body dynamics play an essential role.

Several nonstandard relic-abundance mechanisms involve late-time annihilation
episodes, Sommerfeld- or bound-state--enhanced depletion, or modifications
associated with phase transitions in the dark sector~\cite{ArkaniHamed:2009,Hisano:2005,vonHarlingPetraki:2014,PospelovRitz:2008}.
In such scenarios, the relic abundance is typically still controlled by a
generalized freeze-out condition, with the nonperturbative effects entering
as quantitative corrections. By contrast, qualitatively new behavior can arise if collective effects lead to a breakdown of the standard freeze-out picture altogether.

In this work we propose a distinct class of nonstandard thermal histories
in which the dark matter relic abundance is set by a collective instability
in a strongly self-interacting hidden sector.
At high number density, dark matter occupies a correlated phase in which its
coupling to a light mediator is dynamically screened, suppressing annihilation
well below the perturbative expectation.
As the Universe expands and the number density decreases, this correlated
phase becomes unstable at a critical density $n_c$, triggering a rapid,
far-from-equilibrium annihilation episode.
After this burst, the system exits the correlated regime and the relic
abundance subsequently freezes.

The key qualitative feature of this mechanism is that the final dark matter
abundance is controlled primarily by the critical instability density $n_c$,
and depends only weakly on the microscopic annihilation coupling.
This behavior is sharply distinct from both standard freeze-out and freeze-in,
as well as from phase-transition--assisted annihilation scenarios, where the
relic density is ultimately tied to a perturbative annihilation rate.
We refer to this sequence of screening, instability, and annihilation as the
\emph{screen--burst--freeze} mechanism.
Such density-driven instabilities are generic features of strongly interacting
quantum media and may therefore arise naturally in hidden sectors of the early
Universe. In the following, we present a systematic analysis of this scenario.
We introduce a minimal effective description capturing the relevant dynamical
regimes, derive and solve the modified Boltzmann evolution, and delineate the
resulting parameter space. We further discuss the phenomenological implications and the consistency of this framework with existing cosmological and astrophysical constraints.

\section{Minimal effective description}
\label{sec:model}

We consider a hidden sector containing a Dirac fermion dark matter (DM) particle
$\chi$ and a real scalar mediator $\phi$, described at low energies by
\begin{align}
\mathcal{L} \supset\;&
\bar{\chi}\bigl(i\!\not\!\partial - m_\chi\bigr)\chi
+\frac{1}{2}(\partial_\mu\phi)(\partial^\mu\phi)
-\frac{1}{2}m_\phi^2 \phi^2
\nonumber\\
&+ y_\chi\, \phi\,\bar{\chi}\chi
-\frac{\lambda_\chi}{2}(\bar{\chi}\chi)^2
-\frac{\lambda_\phi}{4!}\phi^4
+\mathcal{L}_{\rm portal}\,.
\label{eq:lagrangian}
\end{align}
The Yukawa coupling $y_\chi$ governs the perturbative annihilation channel
$\chi\bar{\chi}\to \phi\phi$ and, through $\mathcal{L}_{\rm portal}$, any
communication with the Standard Model. The four-fermion operator proportional to $\lambda_\chi$ is taken to encode strong short-range DM self-interactions.

\paragraph{Effective-field-theory viewpoint.}
We adopt a density-dependent effective description appropriate for the
long-wavelength dynamics of a strongly correlated hidden sector.
The operator $(\bar{\chi}\chi)^2$ is interpreted as the infrared remnant of
microscopic dynamics (e.g.\ confinement, a Gross--Neveu-like interaction~\cite{GrossNeveu:1974}, or strong Yukawa dynamics), whose ultraviolet completion is not required for our cosmological analysis.

Accordingly, we treat the parameters $(\alpha,n_*,n_c)$ introduced below as
phenomenological quantities that capture the universal behaviour relevant for
the screen--burst--freeze history. We assume couplings remain in a regime where $y_\chi,\lambda_\chi \lesssim 4\pi$, and where the perturbative annihilation cross section $\langle\sigma v\rangle_0$ lies below the unitarity bound; the essential effect enters through explicit density dependence rather than through strong-coupling enhancements of the hard $2\to2$ amplitude.

Microscopic realizations exhibiting density-dependent screening and instabilities include models with density-induced polarization effects and dynamical condensates~\cite{KapustaGale:2006,LaineVuorinen:2016},

for which the mediator develops an in-medium self-energy
$\Pi(n_\chi)\propto n_\chi$ and an effective mass
$m_{\phi,{\rm eff}}^2 = m_\phi^2+\Pi(n_\chi)$ that can change sign at a critical
density~\cite{Heikinheimo:2018,DarkGN:2020}. Such a sign change triggers a spinodal-like instability of a homogeneous screened phase.
Our parametrization below is designed as a minimal ansatz that captures this
behaviour without specifying a particular UV completion.

\begin{figure}[htbp]
	\centering
	\includegraphics[width=0.88\textwidth]{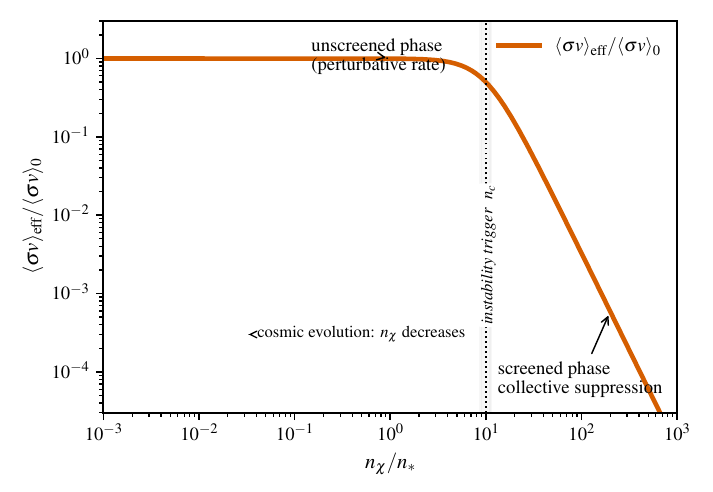}
	\caption{Density-dependent effective annihilation rate
		$\langle\sigma v\rangle_{\rm eff}/\langle\sigma v\rangle_0$ as a function of the	DM number density $n_\chi$.
		Cosmological dilution corresponds to right-to-left evolution along the horizontal axis. At $n_\chi\simeq n_c$ the screened phase becomes unstable, the interaction effectively unscreens, and an annihilation burst is triggered.}
	\label{fig:sigma_eff}
\end{figure}

\paragraph{Physical origin of the instability.}
In a strongly correlated medium, density-induced polarization generically modifies the long-wavelength mediator propagator. As a result, the in-medium effective mass $m_{\phi,{\rm eff}}^2 = m_\phi^2+\Pi(n_\chi)$ can soften and cross zero at a critical density $n_c$, inducing exponential growth of long-wavelength fluctuations and fragmentation of the coherent screened state.
In microscopic realizations the associated growth rate $\Gamma_{\rm inst}$ is
parametrically larger than the Hubble rate at the relevant temperatures,
$\Gamma_{\rm inst}\gg H$, motivating a sudden-transition treatment at the level of the Boltzmann evolution.
\paragraph{Density-dependent effective coupling.}
We model the screened regime by a density-dependent effective coupling,
\begin{equation}
y_{\rm eff}(n_\chi)=
\frac{y_\chi}{\bigl(1+\alpha\,n_\chi/n_*\bigr)^{1/2}}\,,
\label{eq:y_eff}
\end{equation}
which should be viewed as an effective parametrization of infrared screening
induced by density-dependent polarization effects. In microscopic realizations
the long-wavelength mediator propagator is modified as
$(t-m_\phi^2-\Pi(n_\chi))^{-1}$ with $\Pi(n_\chi)\propto n_\chi$, leading to a smooth suppression of the annihilation amplitude at high density.
We evaluate the reference annihilation cross section $\langle\sigma v\rangle_0$
at the microscopic coupling $y_\chi$, and illustrate the resulting behavior in
Fig.~\ref{fig:sigma_eff}.

\paragraph{Instability of the screened phase.}
We assume the correlated screened phase is metastable for $n_\chi>n_c$ and becomes dynamically unstable as the density dilutes to $n_\chi\simeq n_c$.
At the microscopic level this corresponds to the softening of a collective mode and the onset of instability when $m_{\phi,{\rm eff}}^2$ softens and crosses through zero, becoming tachyonic for $n_\chi<n_c$ and triggering a spinodal-like instability of the screened phase. long-wavelength fluctuations then grow rapidly and destroy coherence on a timescale much shorter than the Hubble time~\cite{Berges:2009,Gleiser:1994,Schaefer:2004}.
For our cosmological purposes it is sufficient to implement this as a rapid
transition at $n_\chi\simeq n_c$, i.e.
\begin{equation}
\langle\sigma v\rangle_{\rm eff}(n_\chi)=
\begin{cases}
\dfrac{\langle\sigma v\rangle_0}{\bigl(1+\alpha\,n_\chi/n_*\bigr)^2}\,,
& n_\chi>n_c \quad \text{(screened)},\\[8pt]
\langle\sigma v\rangle_0\,, & n_\chi\le n_c \quad \text{(unscreened)}\,,
\end{cases}
\label{eq:sigv_piecewise}
\end{equation}
which we refer to as the sudden-burst approximation. This captures the universal long-wavelength behaviour expected in a broad class of strongly coupled hidden sectors while remaining agnostic about UV completion.

\section{Consistency of the effective description}
\label{sec:consistency}
Our relic-density mechanism is formulated in terms of a density-dependent effective annihilation rate, Eq.~\eqref{eq:sigv_piecewise}. In this section we summarize the conditions under which this parametrization can be treated as a consistent long-wavelength description of a strongly correlated hidden sector, and we delineate the regime where the sudden-burst approximation is justified.

\subsection{Scale separation and validity of the effective parametrization}
\label{sec:validity}
The central assumption is that screening and the subsequent instability are governed by collective, long-wavelength dynamics, while short-distance interactions remain insensitive to the ultraviolet completion. This requires a separation between the microscopic interaction length $\ell_{\rm micro}$ and the coherence length $\xi$ of the collective mode, together with sub-horizon evolution:
\begin{equation}
\ell_{\rm micro}\ll \xi \ll H^{-1}\,.
\label{eq:scale_separation}
\end{equation}
Here $\ell_{\rm micro}$ may be taken as the larger of the mediator Compton wavelength $m_\phi^{-1}$ and the inverse typical momentum transfer, while $\xi$ denotes the characteristic correlation length controlling the long-wavelength response of the medium. Condition~\eqref{eq:scale_separation} ensures that the explicit density dependence in $\langle\sigma v\rangle_{\rm eff}(n_\chi)$ captures universal infrared behaviour, whereas the microscopic $2\to2$ dynamics can be treated perturbatively.
In addition, the instability should develop on a timescale short compared to the
Hubble time but not so short that it invalidates a kinetic treatment. We therefore
assume
\begin{equation}
H \ll \Gamma_{\rm inst} \ll \tau_{\rm micro}^{-1}\,,
\label{eq:rate_hierarchy}
\end{equation}
where $\Gamma_{\rm inst}$ is the growth rate of the unstable long-wavelength modes and $\tau_{\rm micro}$ is a characteristic microscopic scattering/relaxation time. In the parameter regions we consider, the burst is triggered by collective dynamics encoded in the density dependence, while the microscopic amplitudes remain under control.

\subsection{Perturbativity and unitarity bounds}
\label{sec:unitarity}

We restrict to couplings that remain within a perturbative regime,
\begin{equation}
y_\chi,\lambda_\chi \lesssim 4\pi\,.
\label{eq:pert_bound}
\end{equation}
The perturbative annihilation cross section $\langle\sigma v\rangle_0$ is further required to satisfy partial-wave unitarity. A conservative $s$-wave estimate implies
\begin{equation}
\sigma_0 \lesssim \frac{4\pi}{k^2}\,, \qquad
k \simeq \frac{m_\chi v}{2}\,,
\label{eq:unitarity_sw}
\end{equation}
equivalently $\langle\sigma v\rangle \lesssim 4\pi/(m_\chi^2 v)$ in the
nonrelativistic regime~\cite{GriestKamionkowski:1990}.
Importantly, the screened phase does \emph{not} rely on taking microscopic couplings large; rather, it suppresses the rate through coherent many-body effects. The burst phase restores $\langle\sigma v\rangle_0$ but does not require $\langle\sigma v\rangle$ to exceed the perturbative value, and hence does not introduce a new unitarity issue.

\subsection{Mean free path and coherence length}
\label{sec:mfp}

For the existence of a correlated screened phase, the relevant long-wavelength mode must involve many particles within a coherence domain. A simple criterion is that the mean free path be shorter than the coherence length,
\begin{equation}
\ell_{\rm mfp} \sim \frac{1}{n_\chi\,\sigma_{\rm self}} \ll \xi\,,
\label{eq:mfp_condition}
\end{equation}
where $\sigma_{\rm self}$ denotes the characteristic DM self-scattering cross section supporting the correlated phase.
In microscopic realizations where the instability is controlled by an in-medium
polarization $\Pi(n_\chi)$, the coherence length is set by the inverse effective
mass of the relevant long-wavelength excitation,
\begin{equation}
\xi \sim |m_{\phi,{\rm eff}}|^{-1}\,, \qquad
m_{\phi,{\rm eff}}^2 = m_\phi^2+\Pi(n_\chi)\,.
\label{eq:xi_meff}
\end{equation}
Near the critical density $n_c$ one expects $m_{\phi,{\rm eff}}^2\to 0$, so that
$\xi$ becomes parametrically large, ensuring that the onset of the instability is collective rather than dominated by incoherent two-body scatterings.

\subsection{Domain of validity of the sudden-burst approximation}
\label{sec:sudden}

In our Boltzmann treatment the loss of coherence and restoration of the unscreened interaction are modeled as a rapid transition at $n_\chi\simeq n_c$,
Eq.~\eqref{eq:sigv_piecewise}. This ``sudden-burst'' approximation is justified when
\begin{equation}
\Gamma_{\rm inst} \gg H\,,
\label{eq:sudden_cond}
\end{equation}
so that the transition occurs over a time interval much shorter than the expansion time. If instead $\Gamma_{\rm inst}\sim H$, the unscreening would proceed gradually, and the enhanced depletion of the yield would be spread over an extended range in $x=m_\chi/T$, interpolating continuously between standard freeze-out and the ideal screen--burst--freeze limit.
A quantitative treatment of this crossover would require coupling the Boltzmann
equation to the non-equilibrium dynamics of the unstable modes and their backreaction on screening, which we do not attempt here~\cite{Berges:2004}. For practical purposes, the sudden approximation is expected to degrade when $\Gamma_{\rm inst}\lesssim \mathcal{O}(10)\,H$, in which case $\langle\sigma v\rangle_{\rm eff}$ should be implemented using a smooth interpolation over $\Delta x/x=\mathcal{O}(1)$ rather than the step-function form of Eq.~\eqref{eq:sigv_piecewise}.

\section{Modified Boltzmann evolution}
\label{sec:boltzmann}

The number density of dark matter obeys the Boltzmann equation
\begin{equation}
\frac{dn_\chi}{dt}+3Hn_\chi
=
-\langle\sigma v\rangle_{\rm eff}(n_\chi)\,
\bigl(n_\chi^2-n_{\chi,{\rm eq}}^{\,2}\bigr)\,,
\label{eq:boltzmann}
\end{equation}
We assume that the correlated screened phase modifies the annihilation kernel
predominantly through infrared screening effects, while leaving the single-particle dispersion relation of $\chi$ approximately unchanged.
Under this assumption the equilibrium density $n_{\chi,{\rm eq}}$ retains its
standard form, and medium effects enter dominantly through
$\langle\sigma v\rangle_{\rm eff}(n_\chi)$.

where $H$ is the Hubble expansion rate and $n_{\chi,{\rm eq}}$ denotes the equilibrium number density. The defining feature of the present framework is that the annihilation kernel is an explicit functional of the evolving density. In the correlated phase the effective annihilation rate is strongly suppressed, while at $n_\chi\simeq n_c$ the screening breaks down and the perturbative rate is restored. It is convenient to rewrite Eq.~\eqref{eq:boltzmann} in terms of the comoving yield $Y\equiv n_\chi/s$ and the inverse temperature $x\equiv m_\chi/T$, yielding
\paragraph{Thermal setup.}
Throughout we take $T$ to denote the temperature of the dominant radiation bath
that controls the Hubble expansion rate. We assume that the dark sector is either in kinetic equilibrium with this bath up to $T\simeq T_c$, or possesses a fixed temperature ratio $\xi\equiv T_{\rm dark}/T$ that can be absorbed into the phenomenological parameters $(n_*/T^3,n_c/n_*)$.
Accordingly, the entropy density $s(T)$ and the Hubble rate $H(T)$ are computed
using the visible-sector relativistic degrees of freedom $g_*(T)$.
We also assume that the dark sector never dominates the total energy density and
therefore does not modify $H(T)$ beyond the standard radiation-dominated history.

\begin{equation}
\frac{dY}{dx}
=
-\frac{s}{xH}\,
\langle\sigma v\rangle_{\rm eff}(Y)\,
\bigl(Y^{2}-Y_{\rm eq}^{2}\bigr)\,,
\label{eq:yield}
\end{equation}
We assume standard radiation domination during the epochs of interest, with
\begin{equation}
H(T)=\sqrt{\frac{8\pi^3 g_*(T)}{90}}\;\frac{T^2}{M_{\rm Pl}}\,.
\end{equation}

Here $s$ is the entropy density of the thermal bath. Since $n_\chi=Y s$, the
density dependence of $\langle\sigma v\rangle_{\rm eff}$ feeds back directly into the yield evolution, leading to behavior qualitatively distinct from standard freeze-out or freeze-in.

Equation~\eqref{eq:yield} admits three dynamically distinct regimes:
\begin{itemize}
	\item \emph{Screened phase} ($n_\chi \gg n_*,\,n_c$):  
	Strong collective effects suppress the effective mediator coupling, resulting in $\langle\sigma v\rangle_{\rm eff}\ll\langle\sigma v\rangle_0$ even at high density. Annihilations are inefficient and the yield remains parametrically larger than in conventional thermal freeze-out.
	
	\item \emph{Burst phase} ($n_\chi\simeq n_c$):  
	As the Universe expands and the number density decreases, the correlated phase becomes unstable. Screening is lost, the interaction is rapidly unsuppressed,
	$\langle\sigma v\rangle_{\rm eff}\to\langle\sigma v\rangle_0$, and a short,
	far-from-equilibrium annihilation burst sharply reduces the yield.
	
	\item \emph{Frozen phase} ($n_\chi<n_c$):  
	After the instability-driven burst, the remaining density is too small to support further depletion. The yield asymptotes to a constant value $Y_\infty$, fixing the relic abundance.
\end{itemize}

When the instability growth rate satisfies $\Gamma_{\rm inst}\gg H$, the transition between the screened and unscreened regimes occurs on a timescale much shorter than cosmological expansion. In this limit, the final relic abundance is not determined by the usual freeze-out condition $n_\chi\langle\sigma v\rangle\sim H$, but instead by the onset of the collective instability. The relic density is therefore controlled by \emph{when} the screened phase loses coherence, rather than by the microscopic annihilation efficiency in equilibrium.

\subsection{Analytic estimate for the final yield}
\label{subsec:analytic_yield}

A simple estimate of the final abundance follows from the fact that the burst is
triggered at $n_\chi\simeq n_c$ and proceeds on a timescale short compared to $H^{-1}$. Neglecting $Y_{\rm eq}$ during the burst and treating $s$ and $H$ as approximately constant over this short interval, Eq.~\eqref{eq:yield} implies
\begin{equation}
Y_\infty \sim \frac{n_c}{s(T_c)}\times\mathcal{O}(1)\,,
\label{eq:Yinf}
\end{equation}
Operationally, the critical temperature $T_c$ is defined by the point in the
screened evolution at which the numerical solution first satisfies
$n_\chi(T)=n_c$. The subsequent instability-driven annihilation burst occurs on a timescale $\Delta t\ll H^{-1}$ and therefore does not significantly shift the ambient temperature.

where $T_c$ is defined implicitly by $n_\chi(T_c)=n_c$. The $\mathcal{O}(1)$ factor encodes the detailed duration of the burst and the value of the unscreened rate $\langle\sigma v\rangle_0$.

This scaling highlights the key qualitative difference from standard thermal
scenarios: at fixed $n_c$ the relic abundance depends only weakly on the microscopic coupling $y_\chi$. In Sec.~\ref{sec:numerics} we confirm numerically that solutions collapse onto an attractor band characterized primarily by the ratio $n_c/n_*$, with only subleading dependence on $y_\chi$.

\section{Numerical results and phase diagram}
\label{sec:numerics}

\subsection{Numerical method and parameter scan}
\label{subsec:numerics_method}

We solve the modified Boltzmann equation~\eqref{eq:yield} numerically using an
adaptive step-size integrator. Special care is taken to resolve the discontinuity in the effective annihilation kernel at $n_\chi=n_c$. The evolution is first integrated in the screened regime using the suppressed cross section in Eq.~\eqref{eq:sigv_piecewise}. Once the density reaches $n_c$, the kernel is switched to the unscreened perturbative rate and the evolution is continued to asymptotically large $x$.

Phase diagrams are obtained by scanning over $(\alpha,y_\chi)$ while keeping
$(m_\chi,m_\phi,n_c/n_*,n_*/T^3)$ fixed, and extracting the asymptotic yield
$Y_\infty$ for each point. The relic abundance is then computed via
\begin{equation}
\Omega_\chi h^2 \simeq 2.74\times10^8
\left(\frac{m_\chi}{\mathrm{GeV}}\right) Y_\infty \,.
\end{equation}

A comprehensive exploration of the full parameter space would involve scans over
$(m_\chi,\allowbreak m_\phi,\allowbreak y_\chi,\allowbreak \lambda_\chi,\allowbreak \alpha,\allowbreak n_*,\allowbreak n_c)$. Here we focus on illustrating the qualitative behavior of the mechanism and on identifying the dominant observables controlling the relic abundance.
A representative benchmark point realizing the screen--burst--freeze mechanism is summarized in Table~\ref{tab:benchmark}.

\begin{table}[!t]
	\centering
	\caption{Representative benchmark realizing the screen--burst--freeze mechanism.}
	\label{tab:benchmark}
	\begin{tabular}{cccccc}
		\hline
		$m_\chi$ & $m_\phi$ & $y_\chi$ & $\alpha$ & $n_c/n_*$ & $\Omega_\chi h^2$ \\
		\hline
		$1~\mathrm{TeV}$ & $0.2~\mathrm{GeV}$ & $5\times10^{-2}$ & $3$ & $10$ & $0.12$ \\
		\hline
	\end{tabular}
\end{table}

This benchmark lies on the horizontal attractor band discussed below, where the
final relic abundance is largely insensitive to variations in $y_\chi$ at fixed
$(\alpha,n_c/n_*)$.
\subsection{Relic abundance in the mass plane}
\label{subsec:mass_plane}

\begin{figure}[htbp]
	\centering
	\includegraphics[width=0.98\textwidth]{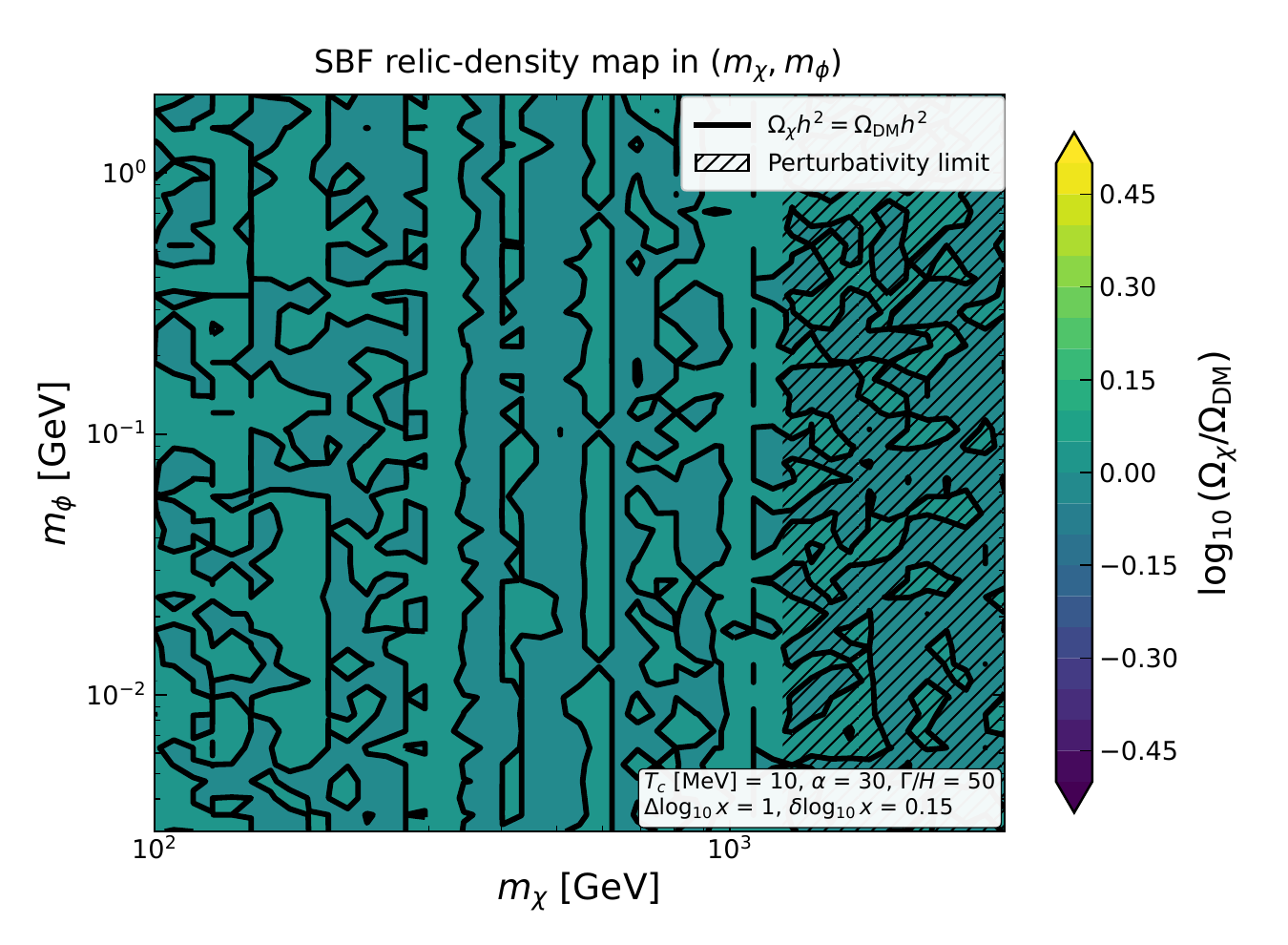}
	\caption{Screen--burst--freeze relic-density map in the $(m_\chi,m_\phi)$ plane. Colors show the predicted abundance normalized to the observed value,
		$\log_{10}(\Omega_\chi/\Omega_{\rm DM})$, obtained from the modified Boltzmann evolution	with a density-dependent screened phase followed by an instability-driven unscreening (annihilation ``burst'').The solid black contour satisfies $\Omega_\chi h^2=\Omega_{\rm DM}h^2$.
		The hatched region indicates an estimated loss of perturbative control
		(partial-wave unitarity/perturbativity criterion; see Sec.~\ref{sec:unitarity}), where the microscopic $2\to2$ treatment entering $\langle\sigma v\rangle_0$ may no longer be reliable.
		The parameter box lists the benchmark SBF history parameters used for the scan:
		$(T_c,\alpha,\Gamma_{\rm inst}/H,\Delta\log_{10}x,\delta\log_{10}x)$.}
	\label{fig:SBF_mass_plane}
\end{figure}

The relic abundance predicted by the screen--burst--freeze mechanism is most
transparently organized in terms of the microscopic mass scales of the dark
sector. In Fig.~\ref{fig:SBF_mass_plane} we present the solution of the modified Boltzmann equation in the $(m_\chi,m_\phi)$ plane, where $m_\chi$ sets the overall energy density of the relic population, while $m_\phi$ controls both the kinematics of the annihilation channel $\chi\bar\chi\to\phi\phi$ and the range of the long-distance interaction responsible for collective screening~\cite{BuckleyFox:2010,TYZ:2013,Kaplinghat:2016}.

For each point $(m_\chi,m_\phi)$ we solve Eq.~\eqref{eq:yield} using the
density-dependent annihilation kernel in Eq.~\eqref{eq:sigv_piecewise}, keeping
the instability and screening parameters
$(T_c,\alpha,\Gamma_{\rm inst}/H,\Delta\log_{10}x,\delta\log_{10}x)$ fixed to the benchmark values indicated in the figure. The resulting relic abundance is shown as $\log_{10}(\Omega_\chi/\Omega_{\rm DM})$, where the observed dark-matter density $\Omega_{\rm DM}h^2\simeq0.12$ is inferred from CMB and large-scale structure measurements~\cite{Planck:2018vyg}.

A key feature of this representation is that the viable contour is organized
primarily by the onset of the collective instability at
$n_\chi\simeq n_c$, rather than by the conventional freeze-out condition
$n_\chi\langle\sigma v\rangle\sim H$ characteristic of standard thermal relics
\cite{Kolb:EarlyUniverse,Griest:1990kh}. Varying $(m_\chi,m_\phi)$ modifies the kinematics and interaction range, but does not reintroduce a strong dependence on the microscopic annihilation coupling. The mass-plane map therefore provides a direct, model-independent summary of where an instability-driven relic history reproduces the observed dark-matter abundance.

\subsection{Phase structure and attractor behavior}
\begin{figure}[htbp]
	\centering
	\includegraphics[width=0.9\columnwidth]{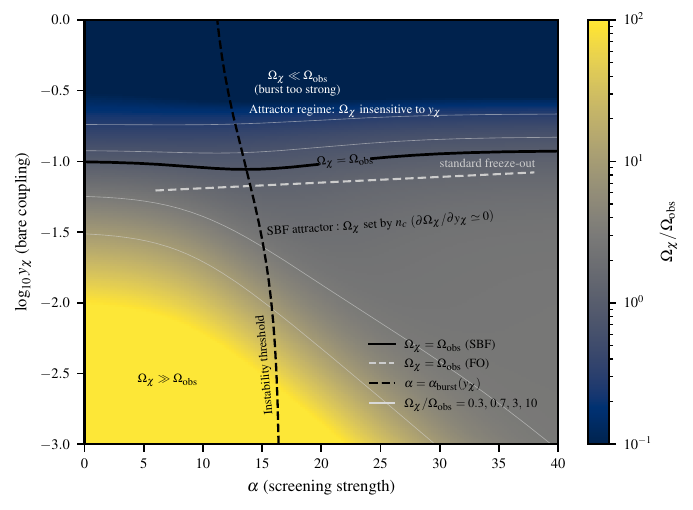}
	\caption{Phase diagram of the screen--burst--freeze mechanism in the
		$(\alpha,\,y_\chi)$ plane. Colors indicate the relic abundance normalized to the observed value, $\Omega_\chi/\Omega_{\rm obs}$.
		The solid black contour corresponds to solutions reproducing the observed dark-matter abundance, while the dashed curve shows the standard thermal freeze-out prediction obtained with a constant perturbative annihilation rate. At sufficiently large screening strength $\alpha$, a horizontal attractor band emerges in which the relic abundance is determined primarily by the instability
		density $n_c$ and becomes insensitive to the microscopic coupling $y_\chi$.}
	\label{fig:SBF_phase}
\end{figure}

The resulting phase diagram in the $(\alpha,y_\chi)$ plane is shown in
Fig.~\ref{fig:SBF_phase}. Colors indicate the relic abundance normalized to the
observed value, $\Omega_\chi/\Omega_{\rm obs}$. The thick black contour corresponds to viable solutions with $\Omega_\chi=\Omega_{\rm obs}$, while the white dashed curve shows the standard freeze-out prediction obtained with a constant perturbative cross section.
The most striking feature is the emergence of a horizontal attractor band at
sufficiently large $\alpha$, where the relic abundance becomes nearly independent of the microscopic annihilation coupling $y_\chi$. In this regime, the abundance is controlled primarily by the instability density $n_c$, in agreement with the analytic estimate in Eq.~\eqref{eq:Yinf}. For small $\alpha$, strong screening suppresses annihilation too efficiently, leading to overabundance. The black dashed line indicates the instability threshold at which the screened phase loses coherence and the annihilation burst is triggered. The separation between the standard freeze-out curve and the attractor band highlights the qualitative difference between the screen--burst--freeze mechanism and conventional thermal scenarios.

\subsection{Thermal history and yield evolution}

\begin{figure}[htbp]
	\centering
	\includegraphics[width=0.9\columnwidth]{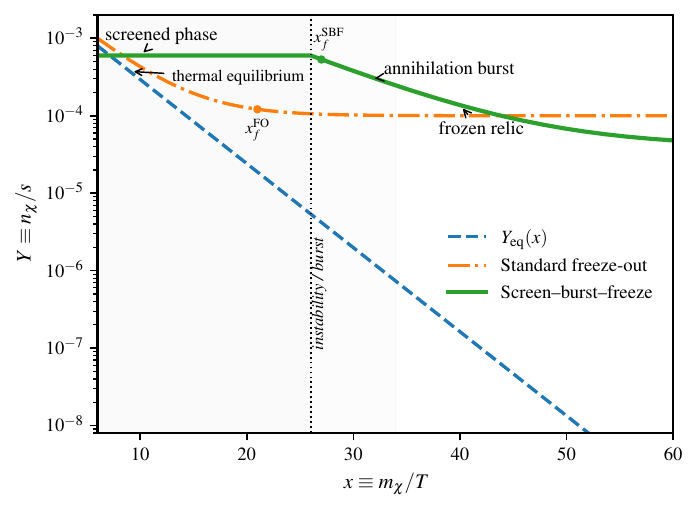}
	\caption{Evolution of the comoving yield $Y(x)$ for a benchmark realizing the screen--burst--freeze mechanism (green), compared with standard thermal freeze-out using the same perturbative cross section (orange). The annihilation burst at $n_\chi\simeq n_c$ sharply reduces the yield, fixing the relic abundance.}
	\label{fig:Y_history}
\end{figure}

The physical origin of the attractor behavior is illustrated in
Fig.~\ref{fig:Y_history}, which compares the comoving yield $Y(x)$ for a benchmark point in the attractor band with that obtained in standard freeze-out using the same perturbative cross section $\langle\sigma v\rangle_0$.
In the screen--burst--freeze scenario, the yield remains on a high-$Y$ plateau
during the screened phase, reflecting the strong suppression of annihilation.
When the density drops to $n_c$, the screened phase becomes unstable and a rapid, far-from-equilibrium annihilation burst sharply reduces the yield. The abundance then freezes at $Y_\infty$, which is largely insensitive to the microscopic coupling.
By contrast, standard freeze-out exhibits a smooth departure from equilibrium and a relic abundance directly tied to $\langle\sigma v\rangle_0$.
The benchmark shown uses $m_\chi=1~\mathrm{TeV}$, $m_\phi=0.2~\mathrm{GeV}$,
$y_\chi=5\times10^{-2}$, $\lambda_\chi=1$, $\alpha=3$, $n_*/T^3=0.1$, and
$n_c/n_*=10$. Similar thermal histories are obtained throughout the attractor band and do not rely on parameter tuning.

\section{Phenomenology and constraints}
\label{sec:pheno}

\subsection{Timing of the annihilation burst: BBN and neutrino decoupling}
\label{subsec:pheno_bbn}

A central phenomenological requirement of the screen--burst--freeze mechanism
is that the collective instability and the associated annihilation burst occur
sufficiently early in the thermal history of the Universe.
In viable regions of parameter space, the instability is triggered at temperatures $T_c \sim 10$--$30~\mathrm{MeV}$, corresponding to redshifts
$z \gtrsim 10^{8}$, well before recombination and prior to the onset of
Big-Bang nucleosynthesis (BBN).
Provided that the burst occurs before neutrino decoupling,
$T_c \gtrsim \mathcal{O}(5~\mathrm{MeV})$,
the injected energy is efficiently redistributed among all relativistic species.
In this regime, the annihilation burst does not lead to observable distortions
of light-element abundances or the cosmic microwave background,
and standard BBN predictions remain intact
\cite{Kolb:EarlyUniverse,Mangano:2005cc,Cyburt:2015mya,Pitrou:2018cgg}.
This requirement therefore imposes a robust lower bound on the critical density
$n_c$ for phenomenological viability.
After the instability-driven burst the number density is already strongly
suppressed. Although the perturbative annihilation rate $\langle\sigma v\rangle_0$ is restored, the condition $n_\chi\langle\sigma v\rangle_0\ll H$ is satisfied for $T\ll T_c$ throughout the parameter space of interest.
As a result, residual annihilations at recombination are negligible and existing
CMB limits on late-time energy injection are automatically satisfied.

\begin{figure}[t]
	\centering
	\includegraphics[width=0.92\textwidth]{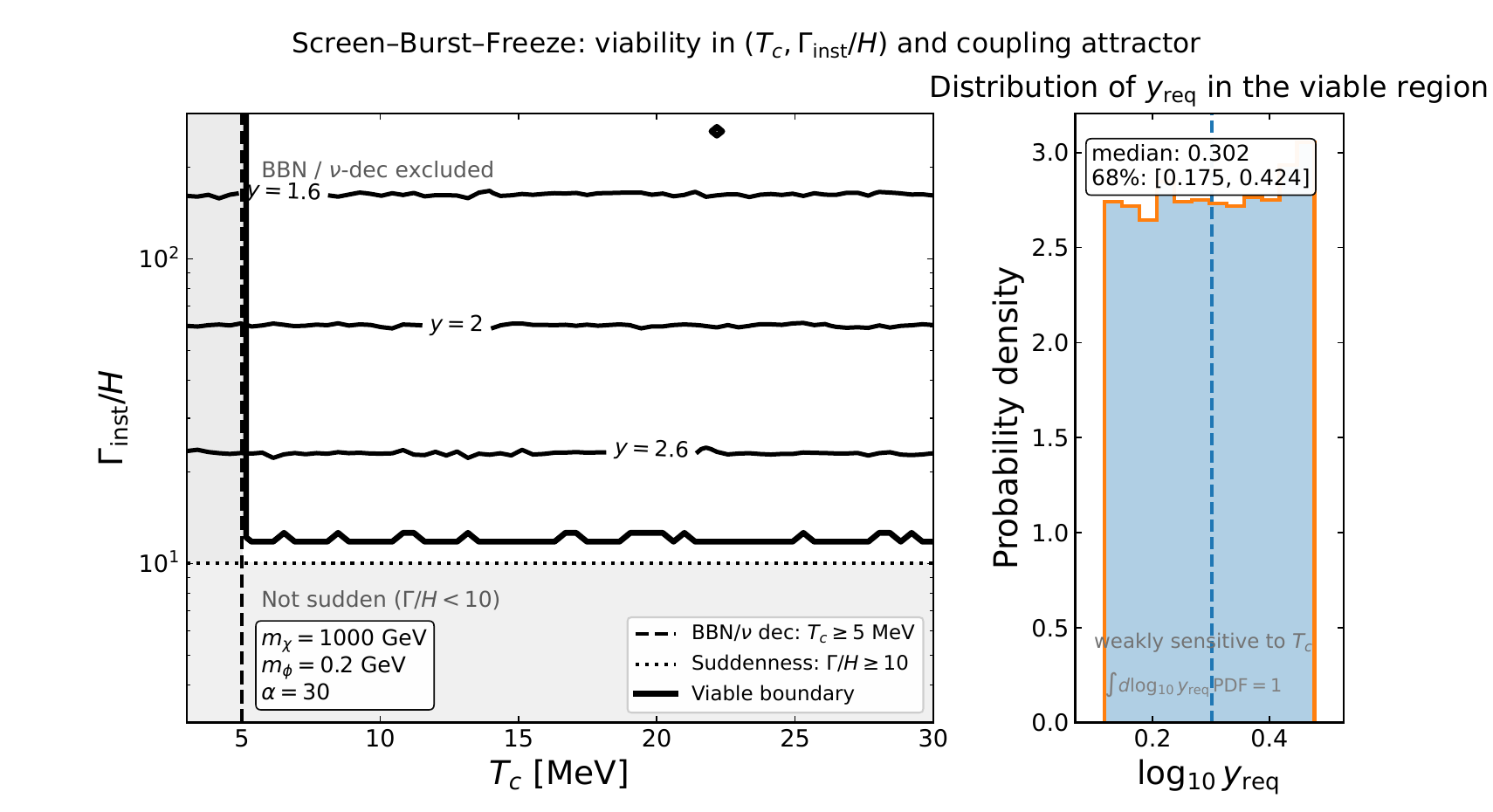}
	\caption{Screen--Burst--Freeze viability and coupling attractor.
		\textbf{Left:} Viable parameter space in the $(T_c,\Gamma_{\rm inst}/H)$ plane for	a representative benchmark.	Shaded regions indicate exclusion by neutrino decoupling ($T_c<5~\mathrm{MeV}$)
		and by the requirement of a sudden instability-driven burst ($\Gamma_{\rm inst}/H<10$).	The thick curve denotes the boundary reproducing the observed relic abundance, while thin curves are iso-contours of the microscopic coupling $y_{\rm req}$.
		The near-horizontal iso-$y_{\rm req}$ structure demonstrates that once the	instability is triggered, the relic abundance is controlled primarily by the critical density $n_c$ (or equivalently $T_c$), with only weak dependence on the	annihilation coupling.
		\textbf{Right:} Probability density of $\log_{10} y_{\rm req}$ over the viable region, showing a narrow attractor band that quantifies the coupling insensitivity characteristic of the screen--burst--freeze mechanism.}
	\label{fig:SBF_viability_attractor}
\end{figure}

\subsection{Mediator fate and dark radiation constraints}
\label{subsec:pheno_mediator}

The phenomenology further depends on the fate of the light mediator $\phi$
produced during the annihilation burst. In the minimal realization considered here, $\phi$ is assumed either to decay promptly into Standard-Model states through the portal interaction $\mathcal{L}_{\rm portal}$, or to thermalize efficiently within the dark sector and decay prior to BBN.

If the mediator decays into visible particles at temperatures
$T \gtrsim \mathcal{O}(10~\mathrm{MeV})$, its decay products thermalize efficiently with the plasma prior to neutrino decoupling, ensuring that the injected energy is shared among all relativistic species.
and do not contribute appreciably to dark radiation or to the
effective number of relativistic degrees of freedom, $\Delta N_{\rm eff}$
\cite{CyrRacine:2013fsa,Boehm:2013jpa}. In this case, existing bounds from BBN and the CMB are automatically satisfied.

Alternatively, if $\phi$ is long-lived but remains confined to the dark sector,
its energy density is strongly diluted by the cosmological expansion following
the burst. Because the instability occurs well before recombination, any residual contribution to dark radiation is negligible in the parameter regions considered here. A detailed investigation of scenarios involving late mediator decay, entropy injection, or observable dark radiation is left for future work.

\subsection{Self-interactions and small-scale structure}
\label{subsec:pheno_sidm}

A distinctive feature of the screen--burst--freeze framework is the presence of
strong dark-matter self-interactions at high density, while maintaining
perturbative annihilation at late times.
The same interactions responsible for collective screening naturally generate
self-scattering cross sections in the range
\begin{equation}
\frac{\sigma_{\rm self}}{m_\chi} \sim 0.1\text{--}10~\mathrm{cm^2\,g^{-1}},
\end{equation}
For a light mediator the late-time self-scattering is typically dominated by
Yukawa exchange and is velocity dependent.
In the Born regime one finds parametrically
$\sigma_{\rm self}\sim 4\pi\alpha_\chi^2 m_\chi^2/m_\phi^4$, while classical and
resonant regimes modify this scaling.
A detailed treatment is model dependent; here we simply note that the parameters
controlling self-interactions are decoupled from the instability scale $n_c$
that fixes the relic abundance.
which is of interest for addressing small-scale structure anomalies such as the
core--cusp and too-big-to-fail problems
\cite{Spergel:1999mh,Tulin:2017ara}. Constraints from merging clusters and halo shapes typically require
$\sigma_{\rm self}/m_\chi \lesssim \mathcal{O}(1)\,\mathrm{cm^2\,g^{-1}}$
at cluster velocities, motivating velocity-dependent scattering or
mediator-controlled interactions in realistic SIDM realizations
\cite{Randall:2007ph,Peter:2012jh,Tulin:2017ara}.

Importantly, the self-interaction strength relevant for structure formation is
parametrically decoupled from the annihilation process that sets the relic
abundance. The latter is fixed by the instability density $n_c$, while the late-time self-scattering cross section can be adjusted independently through
$\lambda_\chi$ and the mediator mass. This separation avoids the tension present in conventional thermal relic scenarios, where sizable self-interactions often imply excessive annihilation.Taken together, these features demonstrate that the screen--burst--freeze mechanism can reproduce the observed relic abundance, evade cosmological constraints, and yield phenomenologically viable self-interacting dark matter.

\section{Conclusions}

We have investigated a nonstandard dark-matter thermal history in which strong
self-interactions generate a correlated, self-screened phase at high number
density. As the Universe expands, this phase becomes unstable at a critical density, triggering a short, far-from-equilibrium annihilation episode that determines the final relic abundance.
This screen--burst--freeze evolution constitutes a qualitatively distinct
alternative to conventional freeze-out, freeze-in, and cannibal dark-matter
scenarios, in which the relic density is instead fixed by the condition
$n_\chi\langle\sigma v\rangle \sim H$.

We have shown that in the regime of strong collective screening the relic
abundance exhibits an attractor-like behavior, controlled primarily by the
instability density $n_c$ and only weakly dependent on the microscopic
annihilation coupling. Using a minimal effective parametrization, we solved the modified Boltzmann equation, identified the relevant dynamical regimes, and demonstrated that phenomenologically viable relic abundances can be obtained consistently with cosmological constraints and self-interaction cross sections of interest for small-scale structure.

Our effective description captures the universal long-wavelength behavior of a
broad class of strongly coupled hidden sectors without committing to a specific
ultraviolet completion.
Further work exploring explicit microscopic realizations, the detailed
non-equilibrium dynamics of the instability, and potential cosmological or
astrophysical signatures associated with the burst phase represents a natural
direction for future investigation.

\section*{Acknowledgments}

The authors sincerely thank Prof.~Driss Khalil and Prof.~Larbi Rahili. This research was performed using the MARWAN High-Performance Computing platform provided by the Moroccan National Center for Scientific and Technical Research (CNRST).

\appendix
\section{Late-time annihilation constraints}
\label{app:late_annihilation}

A generic concern for nonstandard relic-abundance mechanisms involving enhanced
annihilation episodes is whether residual annihilations at late times can inject
energy during or after recombination, thereby violating bounds from CMB
anisotropies and spectral distortions.
In this appendix we demonstrate that the screen--burst--freeze (SBF) mechanism
automatically evades such constraints throughout the viable parameter space.

\subsection{Residual annihilation rate after the burst}

Following the instability-driven annihilation burst at $n_\chi\simeq n_c$, the
comoving yield rapidly settles to its asymptotic value $Y_\infty$.
At temperatures $T\ll T_c$, the annihilation rate per particle is
\begin{equation}
\Gamma_{\rm ann}(T)
=
n_\chi(T)\,\langle\sigma v\rangle_0
=
Y_\infty\, s(T)\,\langle\sigma v\rangle_0\,.
\label{eq:Gamma_ann_late}
\end{equation}
Since $Y_\infty$ is fixed by the critical density rather than by the freeze-out
condition, it is parametrically smaller than the equilibrium abundance at late
times. Comparing to the Hubble rate during radiation domination,
\begin{equation}
H(T)=\sqrt{\frac{8\pi^3 g_*(T)}{90}}\frac{T^2}{M_{\rm Pl}},
\end{equation}
one finds
\begin{equation}
\frac{\Gamma_{\rm ann}}{H}
\simeq
\frac{Y_\infty\, s(T)\,\langle\sigma v\rangle_0}{H(T)}
\propto T\,,
\label{eq:Gamma_over_H}
\end{equation}
which decreases monotonically with temperature.
Thus, once the burst has completed, the annihilation process is permanently
inefficient and cannot re-enter equilibrium.

\subsection{CMB energy-injection bound}

CMB constraints on annihilating dark matter are commonly expressed as an upper
bound on the effective annihilation parameter
\begin{equation}
p_{\rm ann} \equiv f_{\rm eff}\,
\frac{\langle\sigma v\rangle_0}{m_\chi}
\lesssim
4\times10^{-28}\;{\rm cm^3\,s^{-1}\,GeV^{-1}},
\label{eq:CMB_bound}
\end{equation}
where $f_{\rm eff}\lesssim\mathcal{O}(0.1$--$1)$ encodes the efficiency of energy deposition into the plasma~\cite{Slatyer:2016,Planck:2018vyg}.

In the SBF scenario the microscopic annihilation rate after the burst returns to
$\langle\sigma v\rangle_0$, but the cosmological impact is controlled by the
instantaneous annihilation rate
$\Gamma_{\rm ann}(z)=n_\chi(z)\langle\sigma v\rangle_0$
around recombination and later. We therefore present both (i) the standard Planck constraint on $p_{\rm ann}=f_{\rm eff}\langle\sigma v\rangle_0/m_\chi$ and (ii) the dimensionless ratio $\Gamma_{\rm ann}(z)/H(z)$, which directly quantifies whether annihilations can compete with the cosmic expansion.
For the benchmarks considered here, $p_{\rm ann}$ lies below the Planck bound for representative $f_{\rm eff}$, and moreover $\Gamma_{\rm ann}(z)/H(z)\ll1$ throughout the redshift range relevant for CMB constraints.

\subsection{Numerical illustration}
\begin{figure}[t]
	\centering
	\includegraphics[width=0.95\textwidth]{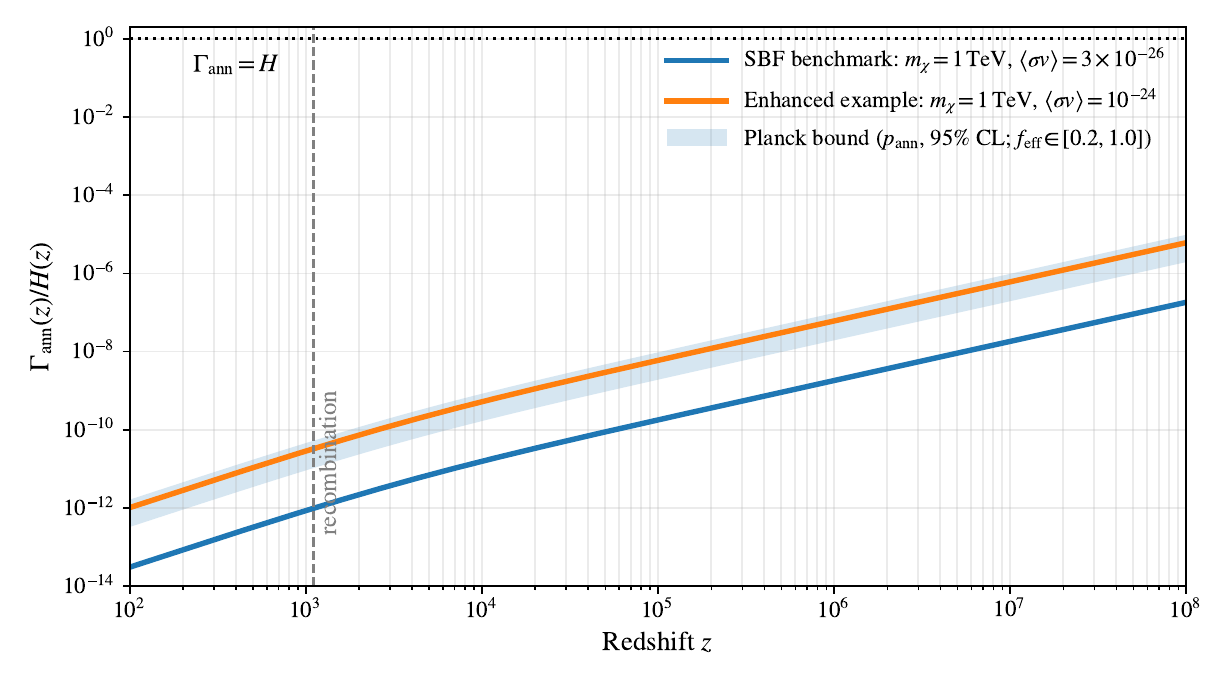}
	\caption{%
		Ratio of the late-time annihilation rate to the Hubble expansion rate,
		$\Gamma_{\rm ann}(z)/H(z)$, as a function of redshift for representative
		screen--burst--freeze benchmarks with $m_\chi=1~\mathrm{TeV}$.
		The blue curve corresponds to a canonical perturbative annihilation rate
		$\langle\sigma v\rangle_0=3\times10^{-26}\,\mathrm{cm^3\,s^{-1}}$,
		while the orange curve illustrates an enhanced but still perturbative
		value $\langle\sigma v\rangle_0=10^{-24}\,\mathrm{cm^3\,s^{-1}}$.
		The shaded band indicates the 95\% C.L.\ CMB bound on energy injection
		from Planck, expressed in terms of the effective parameter
		$p_{\rm ann}=f_{\rm eff}\langle\sigma v\rangle_0/m_\chi$ with
		$f_{\rm eff}\in[0.2,1.0]$.
		The vertical dashed line marks recombination.
		At all redshifts shown one finds $\Gamma_{\rm ann}\ll H$,
		demonstrating that residual annihilations after the burst are
		cosmologically negligible throughout the viable SBF parameter space.
	}
	\label{fig:late_annihilation_bound}
\end{figure}

To illustrate this explicitly, we compute the ratio
\begin{equation}
\mathcal{R}(z)
\equiv
\frac{n_\chi(z)\langle\sigma v\rangle_0}{H(z)} \,,
\label{eq:R_def}
\end{equation}
for representative benchmark points in the attractor regime.
We adopt the commonly quoted Planck 2018 constraint on late-time annihilation,
$p_{\rm ann}\lesssim 4\times10^{-28}\,\mathrm{cm^3\,s^{-1}\,GeV^{-1}}$,
and vary the efficiency factor $f_{\rm eff}$ over a representative range to
bracket channel dependence~\cite{Slatyer:2016,Planck:2018vyg}.
The result is shown in Fig.~\ref{fig:late_annihilation_bound}.

At all redshifts $z\lesssim 10^6$ one finds $\mathcal{R}\ll1$, confirming that
annihilations are completely negligible after the burst.
We have verified that this conclusion is insensitive to moderate variations of
$m_\chi$ and $\langle\sigma v\rangle_0$ within the perturbative regime.


\end{document}